# A Flat Plasmonic Biosensing Interface on Optical Fiber End-Facet via SPP-MIM Hybridization

Chenjia He[1,†], Xiaqing Sun[1,2,†], Hao Zhong[1], Qingfeng Meng[1,5], Xuetong Zhou[3], Sihang Liu[1,4], Li Zheng[3], Xiangyang Kong[2], Shengfu Chen[4], Shengce Tao[5], Tian Yang[1,*]

**Affiliations**

[1]State Key Laboratory of Advanced Optical Communication Systems and Networks, Key Laboratory for Thin Film and Microfabrication of the Ministry of Education, School of Electronic Information and Electrical Engineering, Shanghai Jiao Tong University, Shanghai 200240, China.

[2]School of Material Science and Engineering, Shanghai Jiao Tong University, Shanghai 200240, China.

[3]State Key Laboratory of Materials for Integrated Circuits, Shanghai Institute of Microsystem and Information Technology, Chinese Academy of Sciences, Shanghai 200050, China.

[4]Key Laboratory of Biomass Chemical Engineering, College of Chemical and Biological Engineering, Zhejiang University, Hangzhou 310027, China.

[5]Shanghai Center for Systems Biomedicine, Key Laboratory of Systems Biomedicine (Ministry of Education), Shanghai Jiao Tong University, Shanghai 200240, China

[†]These authors contributed equally to this work.

**Contact Information**

*Correspondence to: tianyang@sjtu.edu.cn, 86-21-34207219.

**Abstract**

We found that the specific dispersion of metal-insulator-metal (MIM) waveguide allows the hybridization of surface plasmon polaritons (SPPs) and the waveguide, which is not possible with dielectric waveguides. The SPP-MIM hybridization structure forms such a meta-film that integrates the previously incompatible respective merits of SPR and LSPR, including flat interfaces, high sensitivities, short evanescent fields and easy coupling with confined light. On the other hand, to achieve stable and reproducible performance is one of the greatest unresolved challenges for the development of nanophotonic biosensors. We point out that the key is to obtain well-controlled biomolecular behaviors using simple physical interfaces, for which the SPP-MIM meta-film provides a capable solution. We embed the SPP-MIM meta-film with a plasmonic crystal cavity and integrate it on a single-mode fiber's end-facet to detect biomolecular interactions. This device demonstrates highly reproducible sensorgrams and convincing detection of biotinylated proteins at down to 30 fM, with the sensorgrams following the Langmuir model. By unprecedentedly having both high sensitivity and high reproducibility, our device proposal provides a comprehensive solution for optical fiber-tip plasmonic devices to turn into a useful industrial biosensing technology.

## 1. Introduction

It has long been expected that integrating plasmonic nanodevices upon optical fiber tips shall provide a disruptive biomolecular sensing technology. For it combines all the advantages of surface plasmon resonance (SPR), fiber-optic communication and the flexible dip-and-read operation mode [1-4]. While industrial SPR sensing technology is well known for high sensitivity and consistency of performance, particularly in the field of biomolecular interaction analysis (BIA), the free-space opto-mechanical system which requires an ultrafine angular resolution, as well as a delicate fluidics system, make the instruments bulky and complicated to operate [5]. At the same time, in the other mainstream BIA industrial technology called biolayer interference (BLI), a miniature glass rod is used to guide light towards the sensing part on the rod's end-facet [6]. Although the multimode glass rod isn't compatible with the fiber-optic communication technology in the general sense, it is still named a fiber-optic sensor. Such a fiber-tip sensing geometry fundamentally improves flexibility, compactness and throughput by using a dip-and-read operation mode. With the successes of SPR and BLI, how to integrate SPR on optical fiber end-facets to combine all of their advantages into a single device stepped into the spotlight [7-17].

In 2016 and 2017, we reported the designs of plasmonic crystal microcavities, as well as a transfer technique that fixed such a microcavity accurately upon the end-facet of a single-mode optical fiber (SMF) [10,11]. The relatively high optical resonance quality-factor (Q) of the microcavity allowed using a fiber-pigtailed super luminescent diode (SLD) to excite SPR efficiently. The limit of detection (LOD) for refractive index reached the $10^{-6}$ RIU level, being two orders of magnitude lower than any other reports about plasmonic devices on SMF tips. In 2022, we reported an upgrade of the fabrication technique to construct a quasi-3D SPR microcavity on the SMF end-facet [15]. The LOD was further lowered to the $10^{-7}$ RIU level, which means its signal-to-noise ratio (SNR) is comparable to the traditional prism-coupled SPR, resolving a key performance limit that comes from the intrinsic difference between planewave coupled SPR and fiber coupled SPR [18,19].

However, having an adequate SNR (or sensitivity) is not yet enough for biomolecular sensing applications, although it was often the sole criterion mentioned in physical device research reports. Another critical aspect is that the interferences from environment or unwanted processes must be eliminated to faithfully extract the signals of the target analytes [20]. From the physical device perspective, this requirement necessitates a simple and controllable physical interface to accommodate the biomolecular binding and dissociation processes. Unfortunately, to date, ultrahigh sensitivity label-free devices, as in a number of reports, often comprise nanostructured interfaces which could result in complicated interfacial processes that haven't been completely understood yet (Fig. 1). For example, one of the possible interference effects is the trapping of gas nanobubbles by nanostructured interfaces (Fig. 1a) [21-23]. Although the trapping mechanism is still under investigation, these nanobubbles are known to exist with high stability [24-26]. It is interesting that the nanobubbles have been observed to both prevent and facilitate biomolecular adsorption [27,28]. Considering that there is a high probability for the nanobubbles to fill in the plasmonic hotspots, they could have remarkable effects on nanoplasmonic devices' sensing performance, including specificity, molecular interactions, analyte transport [29] and consistency. It is worth mentioning there are several critical reviews questioning the reliability of nanoplasmonic biosensing recently [20,30,31].

The same concern is applicable to fiber end-facet SPR devices. That is, the high SNR must be achieved using a simple, controllable and predictable physical interface. The straightforward choice is the flat and uniform surface of a non-structured metallic thin film (Fig. 1c), same as that of the widely proven prism-coupled SPR [32]. However, in all of the existing devices consisting of an SPR microcavity on a SMF end-facet as mentioned above, either penetrating or protruding nanostructures were made in the metallic thin film to couple the fiber-guided lightwaves to the surface plasmon polaritons (SPPs) [10,11,15]. We intensively studied the biosensing performance of these devices and often found the nanostructured interfaces to trap gas bubbles (Fig. 1b), be prone to contamination and have remarkable baseline drifts [15] (See Supplementary Info Section S2). Here we report how to achieve a flat and uniform interface for the SPR microcavities on SMF end-facets, via the hybridization of SPPs and metal-insulator-metal (MIM) gap plasmons. In addition, the devices in this work have a remarkably shorter SPP evanescent depth compared with that of a bare metallic film, so that they are more resistant to the variations of the bulk environment together with an enhanced surface sensitivity. The testing results show great performance improvements compared with previous fiber-tip plasmonic sensors, including less than 0.5 pm min$^{-1}$ baseline drift, detection of 30 fM biotinylated biomolecules, and highly consistent results between different devices and repetitive experiments.

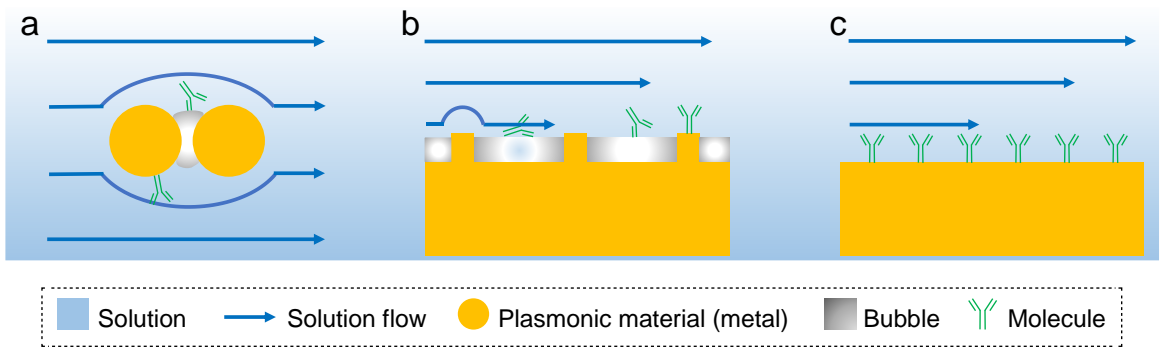

**Fig. 1. Complicated processes at nanostructured interfaces.** Surface gas nanobubbles and distorted mass transfer near nanoparticles (**a**) and a nanostructured surface (**b**), in contrast to a simple and predictable biomolecular adsorption process near a flat surface (**c**).

## 2. Hybridization of SPP and MIM

The major goal of our fiber device design is to achieve efficient coupling between the SMF and the SPPs on a non-structured metallic film. Here, the SPPs are propagating along the water-metal interface, and the SMF is at the film's backside and oriented in the normal direction (Fig. 2a). As we know, the coupling between the illumination and the SPPs can be well achieved if the SMF is replaced by a free-space planewave. This can be achieved either by setting an oblique incidence angle, or by using grating coupling (Fig. 2b) [33]. The coupling efficiency reaches maximum when the phase match condition is satisfied, that is, the SPPs and the incident light having equal wavevector components, $k_x$, along the metallic film's interface.

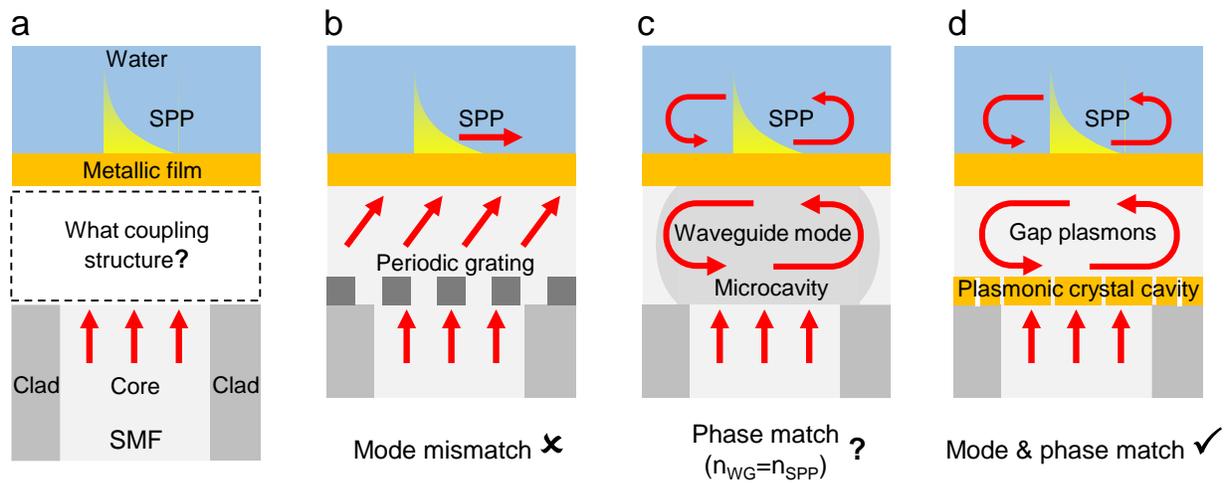

**Fig. 2. Concept of device design.** (**a**) The goal of device design is to excite SPPs on a nonstructured metal film using a SMF from the back. (**b**) The traditional grating coupling structure leads to mode size mismatch and a low coupling efficiency. (**c**) To achieve mode size match, we need a microcavity as an intermediate coupling component between the SMF and the SPPs. This requires phase match between the microcavity waveguide and the SPPs. (**d**) The MIM waveguide can be flexibly tuned to achieve phase match and form an SPP-MIM meta-film. We embed a plasmonic crystal microcavity into the meta-film as our device design.

However, a simple grating coupling scheme won't work for the SMF, due to its limited mode diameter and the considerable numerical aperture. The crucial design concept is to turn the SMF guided lightwave into a microcavity resonance mode, whose propagation phase matches that of the SPPs (Fig. 2c). The phase match condition can be equivalently written as,

$$n_{\text{WG}} = n_{\text{SPP}} \quad (1)$$

where $n_{\text{WG}}$ is the effective refractive index of the microcavity, whose resonance mode is the oscillation of an optical waveguide (WG) mode, and $n_{\text{SPP}}$ is the effective refractive index of the water-side SPPs. Detailed explanation is in the following.

First, as pointed out in refs. [10,18], a microcavity must be formed to confine the SPPs so that a mode size match is obtained between the SPPs and the SMF. Otherwise, the SMF coupled SPR will have a broad and shallow resonance spectrum, resulting in a severely decreased SNR. However, in previous work, the microcavities were built inside the metallic film itself, which results in a nanostructured interface [10,11,15]. In this work, instead, the microcavity is an intermediate component between the metallic film and the SMF, leaving the former non-structured.

Second, it is of crucial importance to achieve phase match between the SPPs and the microcavity so that they can exchange energy efficiently. However, if the microcavity is made of a dielectric waveguide, $n_{\text{WG}}$ would be limited by the dielectric material that comprises the waveguide. From a realistic manufacturing perspective, the waveguide cladding material will have a refractive index significantly larger than that of water. This limitation results in $n_{\text{WG}} > n_{\text{SPP}}$, so that the phase match requirement can't be satisfied. To resolve this problem, we noticed that the gap plasmons guided by the MIM structure can be flexibly tuned by adjusting the gap thickness. In particular, an MIM symmetric waveguide mode can provide a reduced $n_{\text{WG}}$, which eventually turns to negative values as the gap becomes sufficiently thin [34-36]. The device design based on an MIM microcavity is schematically illustrated in Fig. 2d, which satisfies both mode profile matching and phase matching conditions.

In the following part of this section, we will show the calculation results for the non-structured MIM optical waveguide, the hybridization of SPP and MIM, and how the phase matching condition is satisfied. We will leave the discussion about the MIM microcavity and its coupling with the SMF for the next section. Figure 3a shows an infinitely wide MIM waveguide with a water-metal interface, upon which we will focus our calculation here. We call it Insulator-Metal-Insulator-Metal (IMIM) structure, or SPP-MIM meta-film. It consists of an 18 nm thick top gold (Au) film, a 199 nm thick $SiO_2$ gap layer and an infinitely thick Au bottom layer. In this section, for ease of analytical calculation, we adopt the Drude model for Au in which the ohmic loss is neglected.

First, by changing the top Au film to infinitely thick, the photonic band diagram of the MIM waveguide is calculated and plotted in Fig. 3c. It features two distinct branches labelled as *S* and *A*, both being transverse magnetic (TM) waves [36]. Fig. 3g shows the electric field profiles of *S* at three different wavevectors $k_x$, in which *x* is the guided wave's propagation direction. Here the real parts of the complex-numbered electric fields, $E_z$ and $E_x$, are plotted. As shown in the figure, the *S* mode is symmetric with respect to the *x-y* plane that cuts through the center of the $SiO_2$ gap layer, while the *A* mode is anti-symmetric (not plotted). As indicated by the dispersion curves, both modes approach the $SiO_2$-Au interface SPP at extremely large $k_x$, so that their frequencies converge to $\omega_{SP,SiO_2} = \omega_P(1 + \varepsilon_{SiO_2})^{-1/2}$, where $\omega_P$ is the bulk plasmon frequency of Au, and $\varepsilon_{SiO_2}$ is the dielectric constant of $SiO_2$. Notably, while *A* always lies beneath the $SiO_2$ lightline, the *S* dispersion curve splits upward to provide a reduced and tunable $n_{WG}$ to match $n_{SPP}$.

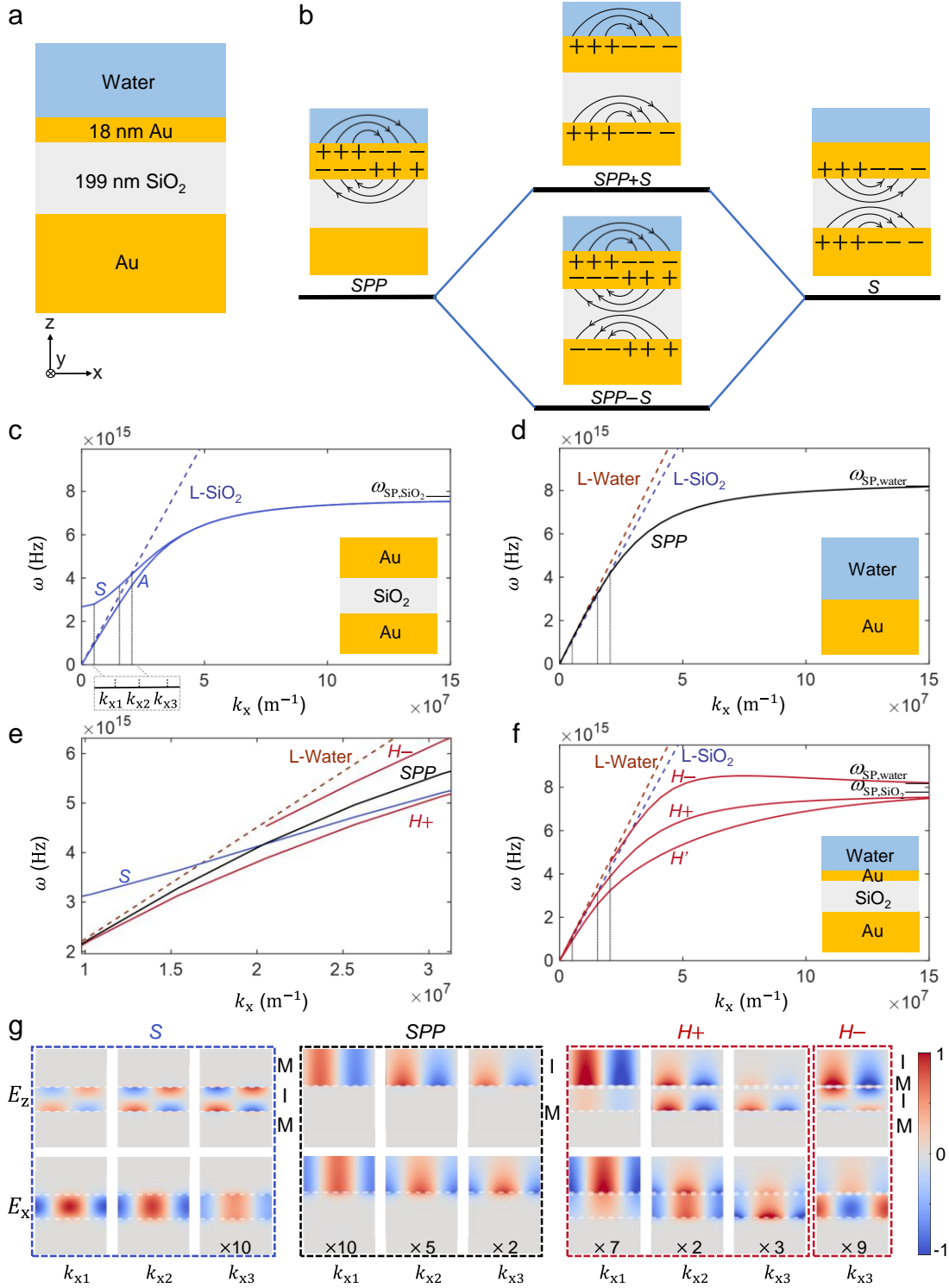

**Fig. 3. Hybridization of MIM and SPP.** (**a**) The IMIM structure includes a water-Au interface that supports SPP propagation and an Au-SiO$_2$-Au MIM waveguide. (**b**) Schematic diagram for *SPP*/*S* hybridization; *S* is a symmetric MIM waveguide mode. The arrows indicate electric fields,

and the signs indicate surface charges. (**c**) The dispersion curves of the MIM waveguide. (**d**) The dispersion curve of the water-Au interface SPP. (**e**) Intersecting between MIM and SPP dispersion curves and anti-crossing after hybridization. (**f**) The dispersion curves of IMIM. The dashed lines in c-f labelled with L are lightlines. (**g**) The real value of electric fields before and after hybridization. Each subfigure includes a period along the *x*-direction and a total height of 840 nm in the *z*-direction. The electric field is normalized to the maximum |***E***| for each case.

Then, the band diagram of SPP at the water-Au interface (with infinitely thick Au) is calculated and plotted in Fig. 3d, which is labelled as *SPP* and converges to $\omega_{\text{SP,water}} = \omega_{\text{P}}(1 + \varepsilon_{\text{water}})^{-1/2}$ at large $k_x$ [37]. Fig. 3g shows its field profiles at the same set of $k_x$'s as above. Of particular interest is where the dispersion curve of *S* intersects that of *SPP*, so that the two can hybridize with each other under the phase matching condition. An enlarged view of this intersection is shown in Fig. 3e. It is worth mentioning that the intersecting point can be flexibly tuned by adjusting the thickness and refractive index of the gap layer [36]. The intersecting frequency can be even higher than $\omega_{\text{SP,SiO}_2}$ as *S* turns into a negative index mode, thus allowing a broad range of working wavelengths.

Lastly, the photonic band diagram of the IMIM structure is calculated and plotted in Fig. 3f. It contains three branches which are labelled *H'*, *H+* and *H−*, respectively. At large $k_x$, their dispersion curves approach those of SPPs at the SiO$_2$-Au interface and the water-Au interface, respectively. An enlarged view (Fig. 3e) shows the anti-crossing of *H+* and *H−* near the intersecting point between *S* and *SPP*, in which *H−* disappears above the water lightline as it becomes a leaky wave. Fig. 3g shows the field profiles of *H+* and *H−* at the same set of $k_x$'s as above. As shown by the figure, the *H+* and *H−* modes predominantly take the form of *SPP*/*S* hybridization near the intersecting point. In the *H+* mode, near the bottom of the gap layer, the $E_z$ component of *SPP* and the $E_z$ component of *S* have the same phase. We call this "*SPP+S*" hybridization (Fig. 3b). Note that the interference between *SPP* and *S* makes the electromagnetic energy localized near the top surface of MIM and the bottom surface of the gap layer. On the contrary, for the *H−* mode, near the bottom of the gap layer, the $E_z$ component of *SPP* and the $E_z$ component of *S* have a $\pi$ phase difference near the intersecting point. We call this "*SPP−S*" hybridization (Fig. 3b). It should be pointed out that the *A* mode also affects the hybridization process and the resulting dispersion curves and field profiles, which will not be discussed here.

Till now, we have successfully designed the *SPP+S* hybridized mode via tuning the MIM optical waveguide to satisfy the phase matching condition, thus forming the SPP-MIM meta-film. This mode has the following desired features for biosensing. First, its dispersion curve lies beneath the water lightline, so that it is always a guided wave. Next, near the intersecting point, its $k_x$ is shifted to be larger than that of *SPP*, which results in a remarkably shorter evanescent depth than *SPP*, as can be seen by comparing the field profiles at $k_{x2}$ (Fig. 3g). As we will discuss later, this decrease in evanescent depth will improve the sensor's resistance to bulk environmental variation and enhance surface sensitivity. At last, its electromagnetic energy distribution is localized near the water-Au interface and near the gap layer's bottom surface, which makes an ideal situation for exciting the water-Au interface SPPs by illuminating the meta-film from the bottom.

## 3. Design of MIM Microcavity on SMF End-Facet

In this section, we report the design of an MIM microcavity by embedding a photonic bandgap structure into the SPP-MIM meta-film, and how it couples the water-Au interface SPPs and the SMF. The device structure is schematically illustrated in Fig. 4a, in which the same IMIM parameters as in Fig. 3a are used, except that the bottom Au layer now has a finite thickness of 65 nm. The bottom Au layer is embedded with a square array of nanoslits which penetrate through it in the $z$ direction.

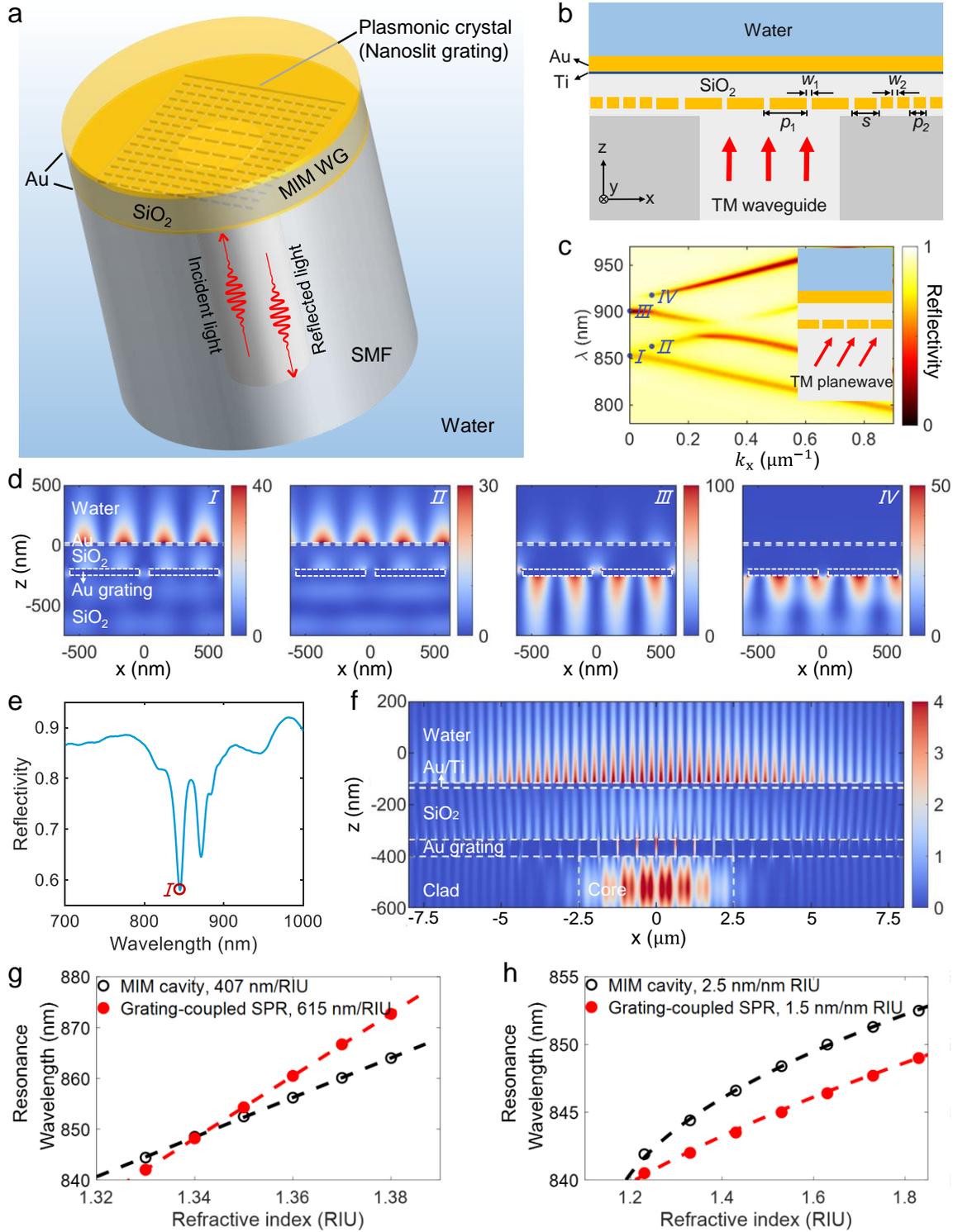

**Fig. 4. Design and simulation of MIM microcavity on SMF end-facet.** (a) Schematic

illustration of the device structure, which consists of a plasmonic crystal cavity in an SPP-MIM meta-film sitting on an SMF end-facet. The Ti adhesion layer is not plotted. (**b**) 2D simulation model (not to scale). $p_1$=622 nm, $w_1$=81 nm, $p_2$=304 nm, $w_2$=123 nm, $s$=400 nm. Top Au thickness 18 nm, Ti adhesion layer thickness 2 nm, $SiO_2$ gap thickness 199 nm, bottom Au thickness 65 nm. (**c**) Photonic band diagram of the SPP-MIM meta-film after being embedded with an infinitely long central nanoslit array, Ti not included. (**d**) Electric field intensity profiles corresponding to Fig. 4c, normalized to the incident light. (**e**) Reflectivity spectrum by using the SMF both to excite the MIM cavity and to collect the reflected light, Ti included. (**f**) Electric field intensity profile corresponding to point *I* in Fig. 4e, normalized to the incident light. (**g**) Bulk refractive index sensitivities of our device and grating-coupled SPR. (**h**) Surface refractive index sensitivities of our device and grating-coupled SPR. The sensitivity values are obtained through linear fitting in Fig. 4g and $2^{nd}$-order polynomial fitting in Fig. 4h, and extracted at the refractive index of 1.33.

Since SPPs, unlike photons, have a single polarization which is TM, it is good enough to consider a two-dimensional simulation model here, as shown in Fig. 4b [10]. In this 2D model, each nanoslit extends infinitely in the *y* direction, and the SMF becomes a 2D waveguide supporting a TM guided mode. The periodically arranged nanoslits in the center are $w_1$=81 nm wide and $p_1$=622 nm apart. They spatially align with the core of the SMF, and form a linear grating to couple the SMF and the *H+* mode. At the same time, the surrounding nanoslits are also periodically arranged, which are $w_2$=123 nm wide and $p_2$=304 nm apart. They form a distributed Bragg reflector to confine the *H+* mode within the microcavity. The number of central nanoslits is seven, and the spacing distance between the central and surrounding nanoslits is $s$=400 nm. These structural parameters have been optimized using the particle swarm algorithm to achieve a maximum value of *coupling efficiency* × *surface sensitivity* × $Q^{1/2}$, near 850 nm optical wavelength. The ohmic loss of Au is included in all of the calculations in this section to better compare with the experimental results.

Moreover, for real devices, an adhesion layer must be added between Au and $SiO_2$. However, typical adhesion metal materials have strong ohmic absorption, which could significantly degrade device performance. For the *H+* mode, its electric field distribution profile barely overlaps with the adhesion metal between the top Au film and the $SiO_2$ gap layer (Fig. 3g). Therefore, it is safe to apply an adhesion metal layer there, for which we use 2 nm thick titanium (Ti). On the other hand, the robust fixing of the bottom Au layer is achieved by filling the nanoslits with $SiO_2$ instead, which will be further discussed as we describe the fabrication procedures. More discussions on the effects of adding adhesion metal layers are included in Supplementary Info Section S3.

To understand how the MIM cavity couples with normal incidence, first, we calculate the photonic band diagram of the SPP-MIM meta-film after it is embedded with the central nanoslit array. This is done by extending the central nanoslit array to an infinite width along the *x* direction, setting a TM planewave incident from the bottom at different angles (Fig. 4c inset), and simulating the reflectivity at different wavelength, λ, and $k_x$ (Fig. 4c). It has been reported that the 2$^{nd}$-order spatial Fourier component of the nanoslit array will produce a photonic bandgap at the center of the First Brillouin Zone [10]. Similarly, in this photonic band diagram, points *I* and *II* correspond to the band edges of the *H+* mode bandgap, points *III* and *IV* correspond to the band edges of another hybridized mode which involves SPPs at the bottom surface of the bottom Au layer. The corresponding $|E^2|$ distribution profiles are shown in Fig. 4d, which have similar features as and whose detailed explanation can be found in our earlier work about nanoslit-array plasmonic crystals [10]. It can be seen that point *I* (*II*) is a bright (dark) mode under normal incidence, and we will use the region near point *I* for biosensing.

The optimized reflectivity spectrum, after including the adhesion metal, is shown in Fig. 4e, using the SMF both to excite the MIM cavity and to collect the reflected light. The spectral dip that corresponds to MIM cavity resonance near bandedge *I* is labelled. It has a spectral depth of ~22% and a Q of ~77. The $|E^2|$ distribution profile of this cavity resonance mode (Fig. 4f) evidently shows that the SMF guided lightwave tunnels through the nanoslits, couples with the MIM gap plasmons, and excites the SPPs on the non-structured top Au film. It is worth mentioning that the optimized device doesn't work right at the *SPP/S* intersecting point, instead, the hybridization proportion can be tuned by moving the intersecting point with respect to the operation wavelength.

In Figs. 4g and 4h, we compare the bulk and surface sensitivities of our device model with those of traditional grating-coupled SPR. The latter is a periodic grating embedded in a thin Au film, which operates under normal illumination with a resonance wavelength near 850 nm (See Supplementary Info Section S4). Since the SPP-MIM meta-film has a significant part of its electromagnetic energy distributed in the $SiO_2$ gap layer (Fig. 4f), rather than in the aqueous solution, its sensitivity to the bulk environment's refractive index change is decreased as compared with that of traditional grating-coupled SPR (Fig. 4g). This means our device has a weaker susceptibility to environmental disturbances, for example, temperature variation. On the other hand, its surface sensitivity is increased (Fig. 4h). Here, surface sensitivity is evaluated by changing the refractive index of a 10 nm thick dielectric layer sitting on the top surface of the top Au film. This high surface sensitivity is attributed to that, for the *H+* mode, the 1/*e*-evanescent depth in the aqueous solution is as short as ~193 nm (Fig. 4f), being significantly shorter than that of traditional grating-coupled SPR which is ~426 nm (Supplementary Fig. S5). At a refractive index of 1.33 (for both bulk and surface layer), the bulk sensitivities of our device and grating-coupled SPR are 407 nm/RIU and 615 nm/RIU, respectively; the surface sensitivities are 2.5 nm/nm RIU and 1.5 nm/nm RIU, respectively. Interestingly, the surface sensitivity of our device shows a nonlinear saturation effect as the surface refractive index increases, which is a distinct feature of LSPR.

## 4. Device Fabrication

Figure 5a shows the flow chart of the device fabrication procedures. First, the MIM layered structure was deposited on a quartz glass substrate, which included electron beam evaporation of the Au and Ti layers, and chemical vapor deposition (CVD) of the $SiO_2$ layer. Then the nanoslit pattern was written into a polymethyl methacrylate (PMMA) resist layer by electron beam lithography. The pattern was transferred into the bottom gold layer by argon ion beam milling. After removing the PMMA resist, a scanning electron microscopy (SEM) image was taken and shown in Fig. 5b. Note that, a square array of nanoslits was made to eliminate polarization dependence. Subsequently, a 160 nm thick $SiO_2$ layer was deposited using atomic layer deposition (ALD) and CVD. The ALD technology ensured sufficient filling of the nanoslits, so that the bottom Au layer was enclosed and tightly fixed by the $SiO_2$ material. This is how we avoid using an adhesion metal layer to fix the bottom Au layer, in order to reduce ohmic loss. Finally, a fiber tip, which had been coated with an UV glue, was aligned with the MIM cavity and placed in contact, and the MIM cavity was stripped off the quartz substrate and transferred onto the fiber end-facet using UV curing. The alignment was done by both microscope observation and monitoring the reflectivity spectrum in real time.

Figure 5c is a top view of the final device under optical microscope. Due to the weak adhesion between the top Au film and the quartz substrate, the MIM cavity had been transferred without noticeable distortion of its pattern. In this figure, the relatively large square-shaped and shaded area is the surrounding nanoslit array, while the relatively small square-shaped and less-shaded area in the center is the central nanoslit array which aligns with the core of SMF. The other features are fabrication marks. For the sensing experiments, each device was mounted on a standard fiber-optic connector, which included a metal ferrule and a ceramic rod (Fig. 5d).

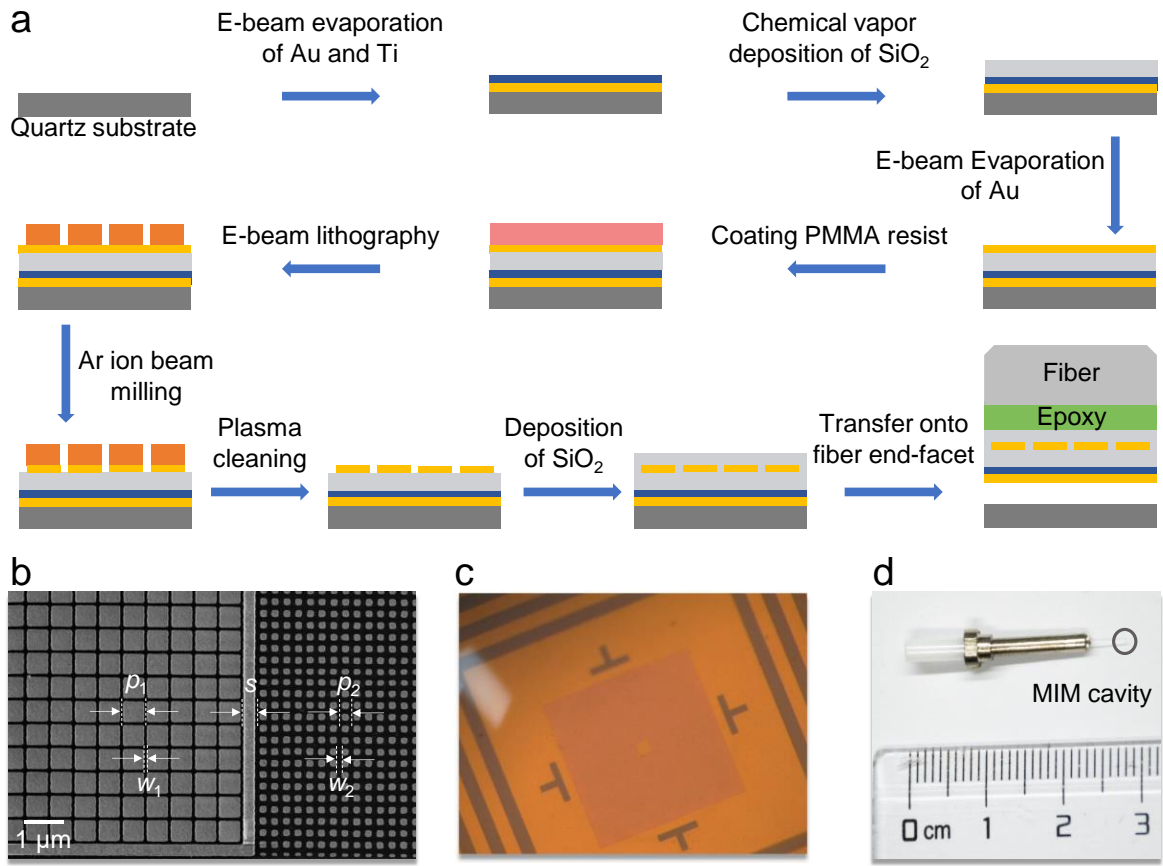

**Fig. 5. Device fabrication.** (**a**) Fabrication procedures. (**b**) SEM image of the plasmonic crystal structure. (**c**) Optical micrograph of the final device on SMF end-facet. (**d**) Packaged device.

## 5. Sensing Experiments

The experiment setup for biomolecular sensing is shown in Fig. 6a. All of the optical fibers used were SMF's. An SLD was used as a broadband light source near 850 nm. A 50%-50% directional coupler both routed the incident light to the sensor probe and routed the reflected light to a CCD spectrometer. An experimental reflectivity spectrum is shown in Fig. 6b.

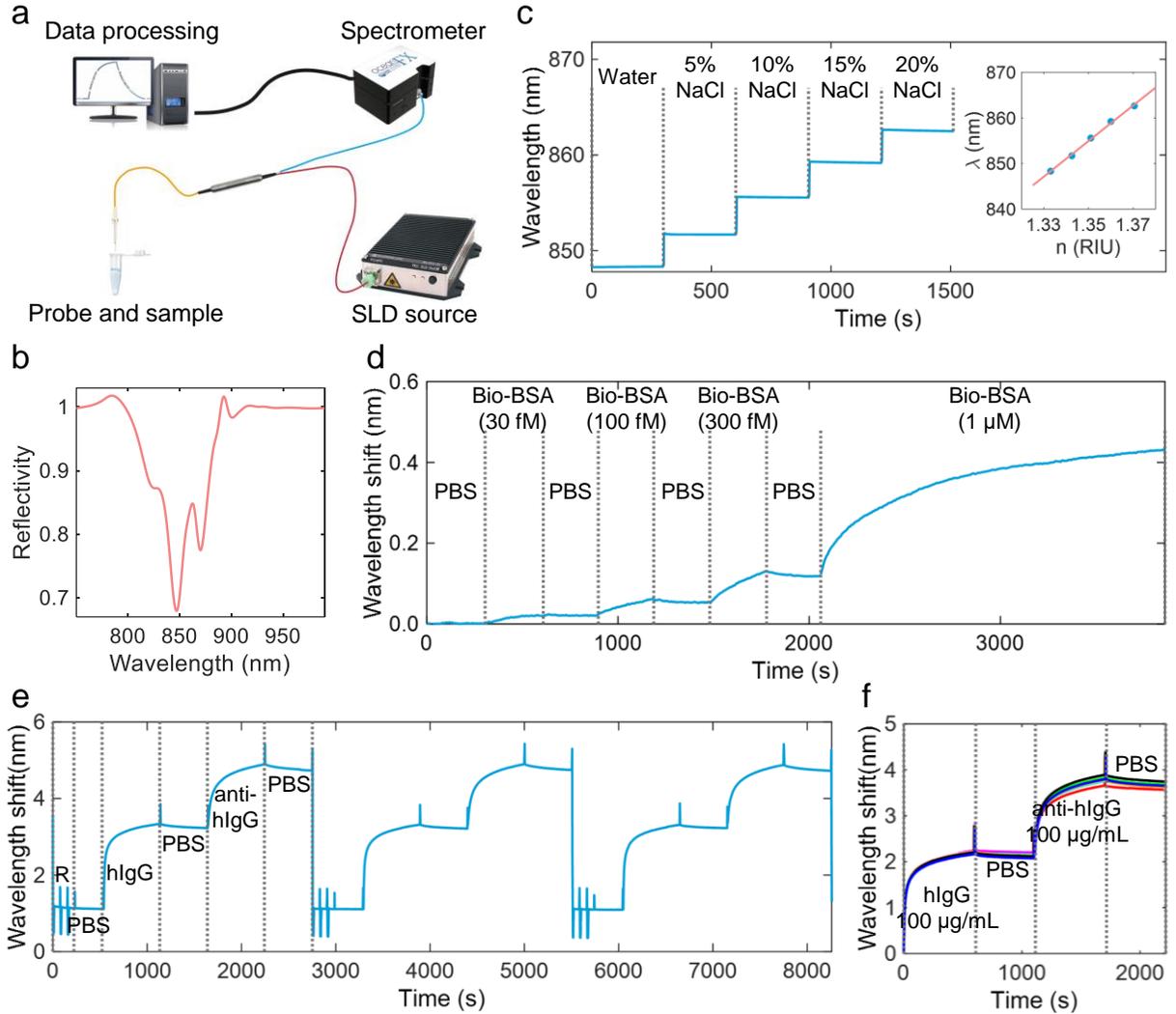

**Fig. 6. Sensing Experiments.** (**a**) Schematic of the experiment setup. (**b**) An experimental reflectivity spectrum in water. (**c**) The MIM cavity's resonance wavelength change as the sensing probe was immersed in different concentrations of NaCl. Inset: linear fitting. (**d**) Real-time binding and dissociation of biotin-BSA onto an SA sensor surface. Baseline drift is corrected. Spikes corresponding to sample switching are removed. (**e**) Real-time binding and dissociation

of hIgG and anti-hIgG, which were repeated by using the regeneration buffer. (**f**) Nine hIgG/anti-hIgG sensorgrams stacked, including three devices and three repetitions per device. PBS: phosphate buffered saline, R: regeneration buffer.

First, we dipped the sensor in different concentrations of NaCl solutions, and recorded the centroid wavelengths of the spectral dip *I* (Fig. 6c). A linear fitting gives a bulk sensitivity of ~401 nm/RIU (Fig. 6c inset), which agrees well with the theoretical prediction.

Then, we measured the interaction between streptavidin (SA) and biotinylated molecules, which is a model system frequently adopted in the literature to demonstrate high sensitivities of various label-free biosensors. Although the exceptionally high affinity of this model system makes it inadequate for drawing a conclusion on BIA experiments in general [30], it serves as a fair comparison with previous claims. The top Au film of the device was first coated with thiol-polyethylene glycol-biotin-streptavidin (HS-PEG-biotin-SA):HS-PEG=1:9, in which PEG serves the purpose of resisting nonspecific binding. Then the device was immersed in biotinylated bovine serum albumin (bio-BSA) solutions, with concentrations ranging from 30 fM to 1 µM. Figure 6d shows the sensorgram which includes multiple real-time binding and dissociation processes. In the figure, the resonance wavelength shift indicates the amount of bio-BSA molecules captured by the SA surface. With an exceptionally high binding rate constant ($k_a$), which is ~$3\times10^9$ $M^{-1}s^{-1}$ as fitted from the sensorgram, this experiment demonstrated detection of bio-BSA at a concentration as low as 30 fM.

It must be noted that achieving a tiny baseline drift is as important as achieving a high SNR for dealing with samples with low concentrations or small molecular weights. In the above experiment, guaranteed by the flat and uniform biosensing interface, the baseline drift in a blank sample was as low as less than 0.5 pm $min^{-1}$, which allowed convincing detection at the lowest concentration. If we consider 2×drift as the limit of detection (LOD), and assume a linear relationship between the resonance wavelength shift and the molecular concentration [20], the LOD for bio-BSA is estimated to be less than 10 fM.

It is important to point out that the sensorgram as obtained here follows the Langmuir model at low analyte concentrations, with some deviations which could be attributed to analyte adsorption on container walls, spatial hindrance and orientation effects of the SA surface, as well as experimental errors. The Langmuir model states that the analyte binding rate is linearly proportional to the analyte concentration. In principle, any kind of sensors should follow the Langmuir model at low molecular concentrations since there is no synergistic effect between different analyte molecules. However, many previous reports that claim detecting analytes at ultralow concentrations did not.

At last, we demonstrated the highly repeatable and consistent performance of each device and between different devices. Here we changed to another model system which is protein A − human immunoglobulin G (hIgG) – anti-hIgG interactions. A layer of protein A was first immobilized on the MIM device's top Au film, then the device was immersed in 100 μg/mL hIgG and 100 μg/mL anti-hIgG sequentially to obtain the binding and dissociation sensorgram, as shown in Fig. 6e. Regeneration of the device was done by washing off the hIgG and anti-hIgG molecules using the Glycine-HCl buffer, so that the experiments were repeated multiple times for each device. The sensorgrams of a total of three devices and three repetitions per device are stacked and compared in Fig. 6f, showing a high level of repeatability and consistency. The individual sensorgrams are shown in Supplementary Info Section S5.

## 6. Discussion

Mounting plasmonic biosensors on optical fiber tips has been an intensively investigated subject for over a decade, for it not only combines the compactness, flexibility and high throughput of fiber-optic techniques and the high sensitivity of the plasmonic evanescent field, but also provides a unique configuration in the form of a miniature probe which is in imminent demand for inspecting ultrasmall volumes of samples and *in vivo* detection with minimum invasion. After the SNR challenge has been overcome for fiber-tip plasmonic devices during 2016-2022 [10,15], these devices have yet to become highly stable and reliable before they are able to compete with the traditional prism-coupled SPR technique and find a broad range of applications.

However, in previous work, the repeatability and reproducibility of biosensing performance have been severely limited by the nanostructured interfaces [31]. There has been a lack of optical designs for efficient excitation of SPPs on a flat and uniform metal-solution interface under normally incident SMF guided lightwaves. To address this challenge, in this work, through the distinct dispersion characteristic and the wide tunability of the symmetric gap plasmon mode in the MIM waveguide, first, we have designed SPP-MIM hybridization; then, by embedding a plasmonic crystal microcavity in the SPP-MIM meta-film, we have realized efficient coupling between the SMF and the SPPs on a flat and uniform metal-solution interface. The stability and repeatability of the devices are remarkably improved compared to previous fiber-tip SPR microcavity devices, based upon which we have demonstrated convincing detection of bio-BSA at down to 30 fM and high uniformity between protein interaction sensorgrams.

From the physical device perspective, to the best of our knowledge, all of the technical obstacles for fiber-tip plasmonic sensors to turn into a successful technique for next generation biomolecular interaction analysis have been overcome at this point. In contrast, the current leading industrial techniques use either SPR [5] or optical cables [6] but not both. At the same time, our work moves the field forward towards the longer-term goal of achieving portable and short turn-around-time medical diagnosis equipment.

Moreover, we show that, through SPP-MIM hybridization, the surface sensitivity is increased and the SPP evanescent depth is significantly decreased. Our further calculation indicates that, by increasing the gap layer's refractive index, the evanescent depth can be reduced to less than 100 nm (near 850 nm optical wavelength) and even higher surface sensitivities can be achieved, which will not be discussed in this report. The harmonious combination of the incompatible merits of propagating SPR and localized SPR (LSPR) makes the SPP-MIM hybridization meta-film a promising candidate to replace traditional SPR and LSPR structures in various scenarios and render new possibilities. By the merits of propagating SPR, we refer to flat sensing interfaces and high SNR (due to high Q) [38]; by the merits of LSPR, we refer to short sensing depths, high resistance to bulk interferences [39,40] and easy coupling with spatially confined light. In this work, we show a first application of this meta-film by successfully coupling a fiber with a flat plasmonic interface on its end-facet.

**Materials and Methods**

SIMULATION METHODS AND SETTINGS

The photonic band diagrams and electromagnetic field distribution profiles of the MIM, SPP and IMIM structures without Au gratings (Fig. 3) were obtained by finite-difference time-domain (FDTD) simulation using the Ansys Lumerical FDTD software. The perfectly matched layer (PML) boundary condition and the Bloch boundary condition were applied for the boundaries along the $z$-direction and the $x$-direction, respectively. Multiple dipole sources were randomly placed and oriented to excite the guided modes, and Fourier transform was performed to obtain the modes. The grid size of the simulation mesh was 0.2 nm in the $z$-direction and 1 nm in the $x$-direction. The dielectric constant of Au was taken to follow the Drude model [41], in which a bulk plasma angular frequency value of $1.37\times10^{16}$ rad/s was used, and the ohmic loss was ignored. The refractive index of water was set to be 1.33, and that of $SiO_2$ was set to be 1.45.

The MIM microcavity structures were also simulated using the same software (Fig. 4). A 2D simulation was performed. The SMF was simulated as a dielectric waveguide with a core width of 5 μm, a core refractive index of 1.45 and a numerical aperture of 0.13. The PML boundary condition was applied for the boundaries along the *z*-direction, and the anti-symmetric boundary condition was applied along the central axis of the device. The grid size of the simulation mesh was from 1 to 5 nm within and near the microcavity, except being 0.2 nm inside and near the Ti adhesion layer. The dielectric constants of Au and Ti were taken from ref. [42].

BIOSENSING EXPERIMENT METHODS

Biomolecular interaction analysis instrument: A homebuilt BIA instrument was used for performing all of the biosensing experiments. A SMF-pigtailed SLD was used as light source, the light was routed to the sensor through a directional coupler, and the reflected light was detected by a CCD spectrometer. The reflectivity spectra were accumulated in the spectrometer at a speed of several hundred spectra per second, and collected by a computer once per second. Then we took the average of the collected spectra and found the centroid wavelength of the SPR dip to be the SPR signal. As shown in Supplementary Fig. S1, the fiber-tip sensor was immersed in an array of centrifuge tubes in sequence, which contained the samples and buffer solutions. The centrifuge tubes' temperature was maintained at 30 °C using proportional-integral-derivative control. The centrifuge tubes were constantly oscillated in the vertical direction at 15 Hz to mitigate the mass transport barrier.

Sensor cleaning and storage: Before each biosensing experiment, the sensors were rinsed with 99.8% ethanol and deionized water. At the end of each experiment, the sensors were immersed in a sucrose solution then air dried to protect the MIM microcavity by covering it with a sucrose droplet. The sucrose droplet was dissolved with water before the next experiment.

Biological and chemical experiment methods: In the bio-BSA sensing experiments, we first coated the sensors' Au surface with a self-assembled layer of thiol-polyethylene-glycol(HS-PEG)-biotin:HS-PEG=1:9, in which PEG was used to prevent nonspecific physical adsorption. Then a layer of SA was immobilized to the biotin end, which was used to capture the bio-BSA analyte. Due to the ultralow analyte concentration, it is important to prevent the analyte from physically adsorbing on the walls of the sample container. Therefore, before the sensing experiments, we used a 10 mg/mL BSA in PBS solution to block the inner surfaces of the centrifuge tubes for 1 h, after which the tubes were rinsed with PBS to remove the blocking solution.

In the hIgG and anti-hIgG sensing experiments, we first coated the sensors' Au surface with a self-assembled layer of HS-PEG-COOH. Then a layer of protein A was immobilized using amine coupling. This was done by immersing the sensors in 0.4 M 1-ethyl-3-(3-dimethylaminopropyl) carbodiimide hydrochloride (EDC) and 0.1 M N-hydroxysulfosuccinimide (Sulfo-NHS) for seven minutes, followed by immersing them in 1 mg/mL protein A solution. Regeneration of the sensors were performed to remove the captured hIgG molecules off the protein A layer. This was done by immersing the sensors in the regeneration buffer, which was Glycine-HCl (10 mM, pH 1.75), for three times and 10 s each time.

**Data availability**

All data that support the findings within the main text and the supplementary information are available from the corresponding author upon reasonable request.

**Acknowledgments**

This work is supported by the National Natural Science Foundation of China (grant No. 62375166, 61975253), the Science and Technology Commission of Shanghai Municipality (grant No. 21N31900200), the National Infrastructures for Translational Medicine (Shanghai) and the Lumieres (Xu Yuan) Biotechnology Company. Device fabrication is supported by the Center for Advanced Electronic Materials and Devices of Shanghai Jiao Tong University. Numerical simulation is supported by the Center for High Performance Computing of Shanghai Jiao Tong University.

**Conflict of interest**

Authors declare no conflicts of interest.

**Supporting Information**

Biosensing experiment instrument; unwanted effects at nanostructured interfaces; effects of adding adhesion metal layers; simulation of traditional grating-coupled SPR; individual protein interaction sensorgrams. (PDF)

**References**

1. G. Kostovski, P. R. Stoddart, A. Mitchell, "The optical fiber tip: an inherently light-coupled microscopic platform for micro- and nanotechnologies," *Adv. Mater.* **26**, 3798-3820 (2014).
2. P. Vaiano, B. Carotenuto, M. Pisco, A. Ricciardi, G. Quero, M. Consales, A. Crescitelli, E. Esposito, A. Cusano, "Lab on fiber technology for biological sensing applications," *Laser Photonics Rev.* **10**, 922-961 (2016).
3. Y. Xiong, F. Xu, "Multifunctional integration on optical fiber tips: challenges and opportunities," *Adv. Photonics* **2**, 064001 (2020).
4. J. Jing, K. Liu, J. Jiang, T. Xu, S. Wang, J. Ma, Z. Zhang, W. Zhang, T. Liu, "Performance improvement approaches for optical fiber SPR sensors and their sensing applications," *Photonics Res.* **10**, 126-147 (2022).
5. https://www.cytivalifesciences.com/en/gb/solutions/protein-research/interaction-analysis-with-biacore-surface-plasmon-resonance-spr.
6. https://www.sartorius.com/en/products/biolayer-interferometry/octet-rh96.

7. E. J. Smythe, M. D. Dickey, J. Bao, G. M. Whitesides, F. Capasso, "Optical antenna arrays on a fiber facet for in situ surface-enhanced Raman scattering detection," *Nano Lett.* **9**, 1132-1138 (2009).
8. G. F. S. Andrade, J. G. Hayashi, M. M. Rahman, W. J. Salcedo, C. M. B. Cordeiro, A. G. Brolo, "Surface-enhanced resonance Raman scattering (SERRS) using Au nanohole arrays on optical fiber tips," *Plasmonics* **8**, 1113-1121 (2013).
9. P. Jia, Z. Yang, J. Yang, H. Ebendorff-Heidepriem, "Quasiperiodic nanohole arrays on optical fibers as plasmonic sensors: fabrication and sensitivity determination," *ACS Sens.* **1**, 1078-1083 (2016).
10. X. He, H. Yi, J. Long, X. Zhou, J. Yang, T. Yang, "Plasmonic crystal cavity on single-mode optical fiber end facet for label-free biosensing," *Appl. Phys. Lett.* **108**, 231105 (2016).
11. Z. Lei, X. Zhou, J. Yang, X. He, Y. Wang, T. Yang, "Second-order distributed-feedback surface plasmon resonator for single-mode fiber end-facet biosensing," *Appl. Phys. Lett.* **110**, 171107 (2017).
12. M. Principe, M. Consales, A. Micco, A. Crescitelli, G. Castaldi, E. Esposito, V. La Ferrara, A. Cutolo, V. Galdi, A. Cusano, "Optical fiber meta-tips," *Light-Sci. Appl.* **6**, e16226 (2017).
13. J. A. Kim, D. J. Wales, A. J. Thompson, G. Z. Yang, "Fiber-optic SERS probes fabricated using two-photon polymerization for rapid detection of bacteria," *Adv. Opt. Mater.* **8**, 1901934 (2020).
14. B. Du, D. Yang, Y. Ruan, P. Jia, H. Ebendorff-Heidepriem, "Compact plasmonic fiber tip for sensitive and fast humidity and human breath monitoring," *Opt. Lett.* **45**, 985-988 (2020).
15. X. Sun, Z. Lei, H. Zhong, C. He, S. Liu, Q. Meng, Q. Liu, S. Chen, X. Kong, T. Yang, "A quasi-3D Fano resonance cavity on optical fiber end-facet for high signal-to-noise ratio dip-and-read surface plasmon sensing," *Light: Adv. Manuf.* **3**, 46 (2022).
16. H.-T. Kim, M. Yu, "On-fiber multiparameter sensor based on guided-wave surface plasmon resonances," *J. Lightwave Technol.* **40**, 2157-2165 (2022).
17. F. Wang, X. Li, S. Wang, Y. Cao, L. Zhang, Y. Zhao, X. Dong, M. Zheng, H. Liu, W. Lu, X. Lu, C. Huang, "3D fiber-probe surface plasmon resonance microsensor towards small volume sensing," *Sens. Actuat. B-Chem.* **384**, 133647 (2023).
18. T. Yang, X. He, X. Zhou, Z. Lei, Y. Wang, J. Yang, D. Cai, S.-L. Chen, X. Wang, "Surface plasmon cavities on optical fiber end-facets for biomolecule and ultrasound detection," *Opt.*

*Laser Technol.* **101**, 468-478 (2018).

19. X. Fan, "Sensitive surface plasmon resonance label-free biosensor on a fiber end-facet," *Light-Sci. Appl.* **11**, 325 (2022).

20. T. Yang, S. Chen, X. He, H. Guo, X. Sun, "How to convincingly measure low concentration samples with optical label-free biosensors," *Sens. Actuat. B-Chem.* **306**, 127568 (2020).

21. X. Zhang, D. Lohse, "Perspectives on surface nanobubbles," *Biomicrofluidics* **8**, 041301 (2014).

22. M. Alheshibri, J. Qian, M. Jehannin, V. S. J. Craig, "A history of nanobubbles," *Langmuir* **32**, 11086-11100 (2016).

23. Y. Wang, X. Li, S. Ren, H. Tedros Alem, L. Yang, D. Lohse, "Entrapment of interfacial nanobubbles on nano-structured surfaces," *Soft Matter* **13**, 5381-5388 (2017).

24. D. Lohse, X. Zhang, "Surface nanobubbles and nanodroplets," *Rev. Mod. Phys.* **87**, 981-1035 (2015).

25. B. H. Tan, H. An, C.-D. Ohl, "Surface nanobubbles are stabilized by hydrophobic attraction," *Phys. Rev. Lett.* **120**, 164502 (2018).

26. L. M. Zhou, X. Y. Wang, H. J. Shin, J. Wang, R. Z. Tai, X. H. Zhang, H. P. Fang, W. Xiao, L. Wang, C. L. Wang, X. Y. Gao, J. Hu, L. J. Zhang, "Ultrahigh density of gas molecules confined in surface nanobubbles in ambient water," *J. Am. Chem. Soc.* **142**, 5583-5593 (2020).

27. Z. Wu, X. Zhang, X. Zhang, J. Sun, Y. Dong, J. Hu, "In situ AFM observation of BSA adsorption on HOPG with nanobubble," *Chin. Sci. Bull.* **52**, 1913-1919 (2007).

28. H. Seo, N. Jung, D. Lee, S. Jeon, "Influence of nanobubbles on molecular binding on gold surfaces," *Colloids and Surfaces A: Physicochemical and Engineering Aspects* **336**, 99-103 (2009).

29. B. Špačková, N. S. Lynn, J. Slabý, H. Šípová, J. Homola, "A route to superior performance of a nanoplasmonic biosensor: consideration of both photonic and mass transport aspects," *ACS Photonics* **5**, 1019-1025 (2018).

30. A. Dahlin, "Biochemical sensing with nanoplasmonic architectures: we know how but do we know why?" *Annu. Rev. Anal. Chem.* **14**, 281-297 (2021).

31. H. Altug, S.-H. Oh, S. A. Maier, J. Homola, "Advances and applications of nanophotonic biosensors," *Nat. Nanotechnol.* **17**, 5-16 (2022).

32. J. Homola, "Surface plasmon resonance sensors for detection of chemical and biological

species," *Chem. Rev.* **108**, 462-493 (2008).

33. J. Homola, I. Koudela, S. S. Yee, "Surface plasmon resonance sensors based on diffraction gratings and prism couplers: sensitivity comparison," *Sens. Actuat. B-Chem.* **54**, 16-24 (1999).
34. H. Shin, S. Fan, "All-angle negative refraction for surface plasmon waves using a metal-dielectric-metal structure," *Phys. Rev. Lett.* **96**, 073907 (2006).
35. H. J. Lezec, J. A. Dionne, H. A. Atwater, "Negative refraction at visible frequencies," *Science* **316**, 430-432 (2007).
36. T. Yang, K. B. Crozier, "Analysis of surface plasmon waves in metal-dielectric-metal structures and the criterion for negative refractive index," *Opt. Express* **17**, 1136-1143 (2009).
37. S. A. Maier, "*Plasmonics: fundamentals and applications*," Chapter 2 (Springer, 2007).
38. J. Slabý, J. Homola, "Performance of label-free optical biosensors: what is figure of merit (not) telling us?" *Biosens. Bioelectron.* **212**, 114426 (2022).
39. M. Svedendahl, S. Chen, A. Dmitriev, M. Käll, "Refractometric sensing using propagating versus localized surface plasmons: a direct comparison," *Nano Lett.* **9**, 4428-4433 (2009).
40. S. J. Zalyubovskiy, M. Bogdanova, A. Deinega, Y. Lozovik, A. D. Pris, K. H. An, W. P. Hall, R. A. Potyrailo, "Theoretical limit of localized surface plasmon resonance sensitivity to local refractive index change and its comparison to conventional surface plasmon resonance sensor," *J. Opt. Soc. Am. A* **29**, 994-1002 (2012).
41. A. D. Rakić, A. B. Djurišić, J. M. Elazar, M. L. Majewski, "Optical properties of metallic films for vertical-cavity optoelectronic devices," *Appl. Opt.* **37**, 5271-5283 (1998).
42. E. D. Palik, "*Handbook of optical constants of solids*," (Academic Press, 1998).

Supplementary Information for

# A Flat Plasmonic Biosensing Interface on Optical Fiber End-Facet via SPP-MIM Hybridization

Chenjia He, Xiaqing Sun, Hao Zhong, Qingfeng Meng, Xuetong Zhou, Sihang Liu, Li Zheng, Xiangyang Kong, Shengfu Chen, Shengce Tao, Tian Yang*

*Correspondence to: tianyang@sjtu.edu.cn

**This file includes:**

**Sections**
S1. Biosensing experiment instrument
S2. Unwanted effects at nanostructured interfaces
S3. Effects of adding adhesion metal layers
S4. Simulation of traditional grating-coupled SPR
S5. Individual protein interaction sensorgrams

**References** S1-S5

**Figures** S1-S6

## S1. Biosensing experiment instrument

A homebuilt biomolecular interaction analysis instrument was used for performing all of the biosensing experiments, as described in Materials and Methods and shown in Fig. S1.

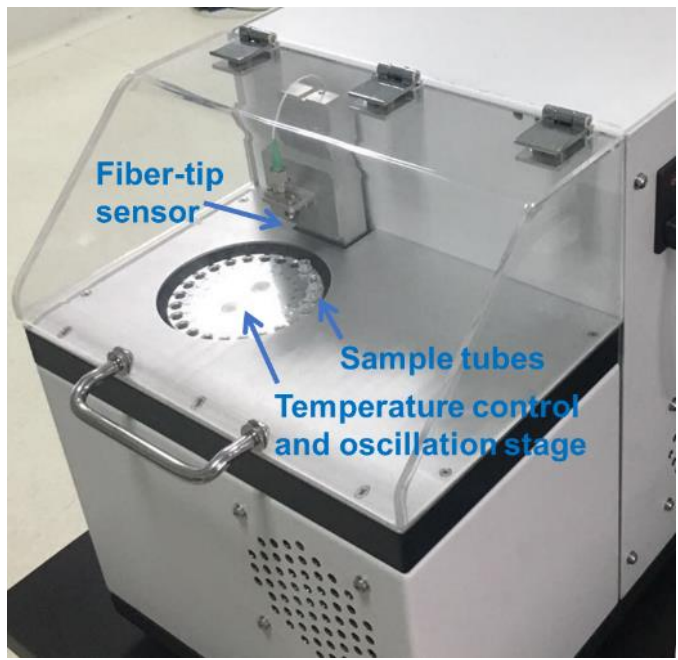

**Fig. S1. Biomolecular interaction analysis instrument.**

## S2. Unwanted effects at nanostructured interfaces

### S2.1 Surface gas nanobubbles

In recent years, a number of experimental and theoretical efforts have been reported to find out about how gaseous nanobubbles are stably formed at the interfaces between water and solid surfaces under hydrophobic forces [24,25] and nucleate at nanostructures [S1-S3]. It is commonly agreed that these nanobubbles come from the dissolved gases and can only exist stably at very small sizes. They could have considerable effects on a variety of interfacial processes such as mass and charge transfer [S4], molecular interactions [27,28] and drag reduction [S5].

In Fig. S2, we show the formation of surface nanobubbles on a waffle-patterned Au surface. As shown in Fig. S2a and S2b, this device contained an array of protruding Au nanostrips which divided a flat Au surface into several hundred nanometer wide square pits. The nanostrips were 50 nm wide and less than 20 nm tall. A similar surface was used in ref. [15] to achieve a high SNR for SMF end-facet SPR sensors. In Fig. S2b, we show the surface topography obtained using atomic force microscopy (AFM), for when the device was immersed in water. In Fig. S3b, we compare the line scanning AFM results for when the device was immersed in water and when it was dry. The comparison indicates a ~ 7 nm thick gas layer in the former. This example illustrates that even a tiny proportion of nanostructures on a flat solid surface can significantly change the surface properties by inducing surface nanobubbles. Though it is a common practice to prepare hydrophilic surfaces for biosensing applications, the process of coating the hydrophilic material can be significantly affected by the surface nanobubbles.

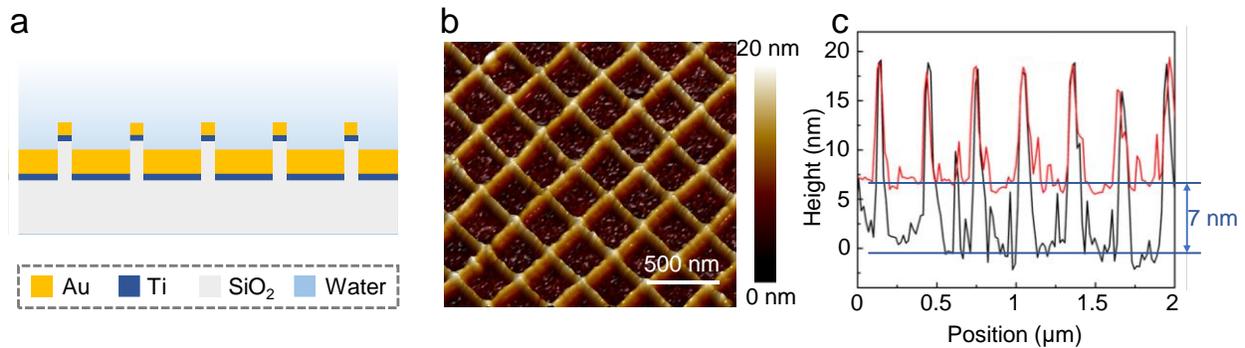

**Fig. S2. Gaseous nanobubbles at nanostructured interfaces.** (**a**) Schematic illustration of a waffle-patterned Au surface. (**b**) AFM image of the device immersed in water. (**c**) AFM line scanning of the device in water (red) and dry (black).

*S2.2 Surface contamination*

We also found that the nanostructured Au surface with a waffle-like pattern, as in Fig. S2, appeared significantly contaminated after being immersed in biomolecule solutions. In Fig S3, we compare it with the MIM microcavity device to show the contamination effect. First, we fabricated both structures on SMF end-facets. Then we cleaned them using 99.8% ethanol and deionized water. After being dried under nitrogen blow, their initial optical micrographs are shown in Fig. S3a and S3c. Then we immersed them in BSA solutions (in PBS buffer) from 10 ng/mL to 100 µg/mL; they were kept in each concentration of BSA solution for ten minutes, with tenfold increase in concentration in each step (See Materials and Methods and ref. [15] for experiment details). At last, they were cleaned again using 99.8% ethanol and deionized water. The optical micrographs of the devices after being dried under nitrogen blow are shown in Fig. S3b and S3d. It is apparent that the nanostructured Au surface was left significantly contaminated. At the same time, the MIM microcavity with a flat surface didn't look noticeably different from the initial device, indicating the physically adsorbed BSA layer on this device was thin and uniform.

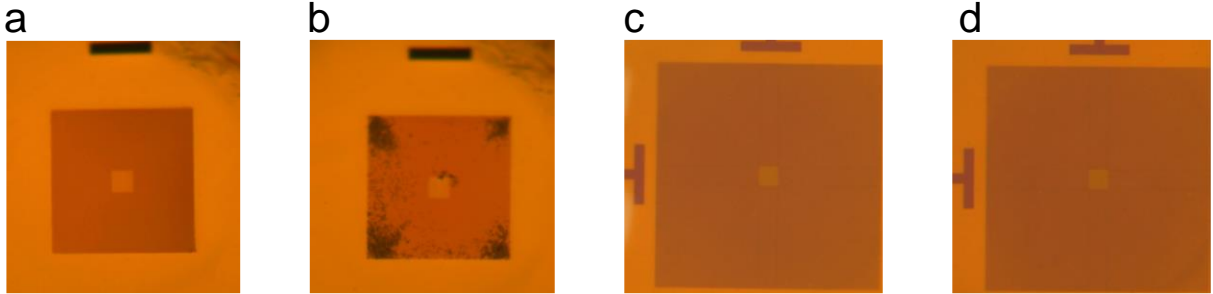

**Fig. S3. Physical adsorption of biomolecules on nanostructured and flat Au surfaces.** Optical micrographs of a waffle-patterned Au surface (**a**) before and (**b**) after being immersed in BSA solutions, in comparison with optical micrographs of an MIM microcavity device with a flat Au surface (**c**) before and (**d**) after being immersed in BSA solutions.

## S3. Effects of adding adhesion metal layers

In Fig. S4, we compare the reflectivity spectra when the 2 nm thick Ti adhesion layer is added to different positions of the device. Otherwise, the device structure and parameters are the same as in the paper. It can be seen that, compared with no Ti layer, when the Ti layer is added between the top Au layer and the $SiO_2$ gap layer, the depth and Q of the SPR spectral dip are only minimally degraded. This is due to that the *H+* mode's electric field intensity is relatively weak at this position (Fig. 4f). However, when the Ti layer is added between the bottom Au grating and the $SiO_2$ gap layer, the depth and Q of the SPR spectral dip are significantly degraded. This is due to that the *H+* mode's electric field intensity is much stronger at this position (Fig. 4f). Therefore, it is not much of a problem to add a metal adhesion layer above the gap layer in the IMIM structure; at the same time, we had better avoid adding a metal adhesion layer beneath the gap layer. The reflectivity spectra were obtained using an FDTD 2D simulation model as mentioned in the paper.

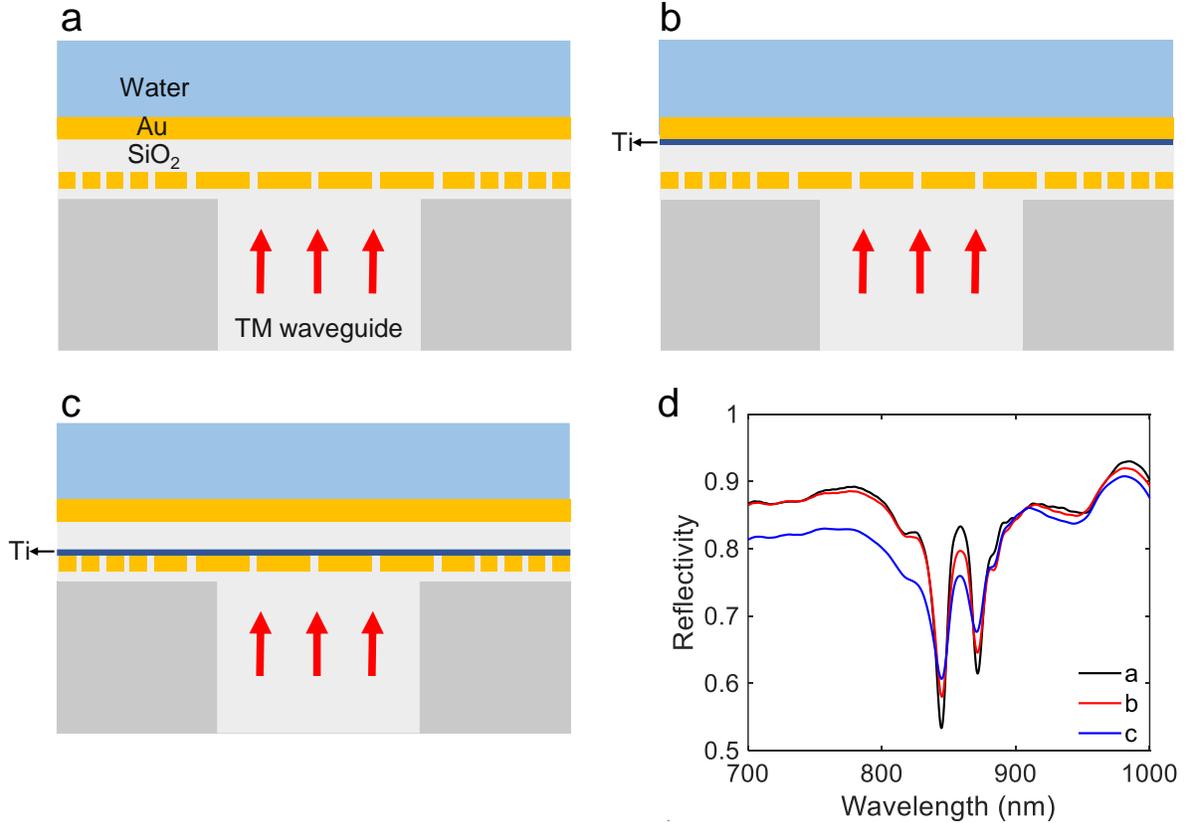

**Fig. S4. Effects of the Ti adhesion layer.** Schematic illustration of two-dimensional simulation models for MIM microcavity on SMF end-facet, with (**a**) no Ti layer, (**b**) a 2 nm Ti layer above the gap layer, and (**c**) a 2 nm Ti layer beneath the gap layer. (**d**) The simulated reflectivity spectra.

## S4. Simulation of traditional grating-coupled SPR

A traditional grating-coupled SPR structure is plotted in Fig. S5a. It is a periodic nanoslit grating in a 40 nm thick Au thin film, which sits on a $SiO_2$ substrate. The period of the grating is 630 nm, the nanoslit width is 50 nm, and the length of the grating is infinite. Under normally incident TM illumination, it shows an SPR spectral dip at 842 nm in the reflectivity spectrum (Fig. S5b). Its electric field intensity profile on SPR resonance shows an evanescent depth of 426 nm in water (Fig. S5c). Its bulk and surface sensitivities were calculated and plotted in Fig. 4g and 4h.

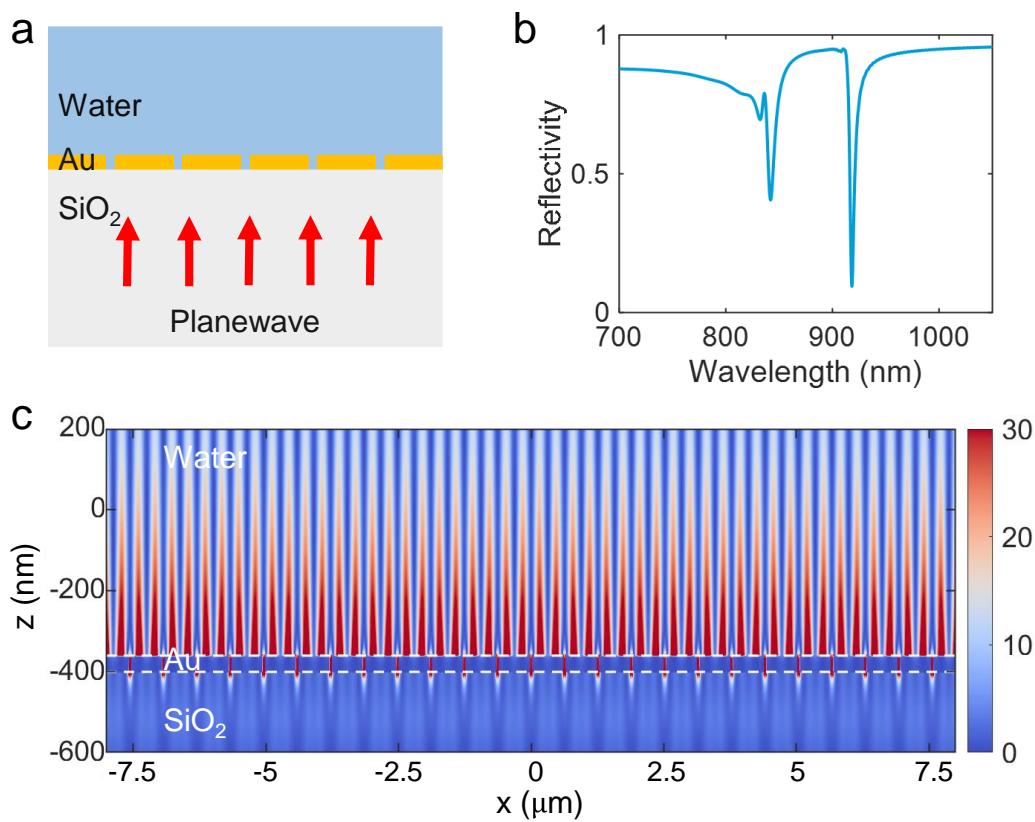

**Fig. S5. Grating-coupled SPR.** (**a**) Schematic illustration of a nanoslit grating in a 40 nm thick Au film. Grating period is 630 nm, slit width is 50 nm. (**b**) FDTD simulation result of the reflectivity spectrum. (**c**) FDTD simulation result of the electric field intensity distribution at 842 nm.

## S5. Individual protein interaction sensorgrams

The protein A − hIgG − anti-hIgG biomolecular interaction experiments were performed with three different sensor devices. For each device, we repeated the experiments for three times by using the regeneration buffer to remove the hIgG molecules off the protein A layer. The sensorgrams for all of the nine experiments are shown in Fig. S6, which are stacked in Fig. 6f to show a good repeatability.

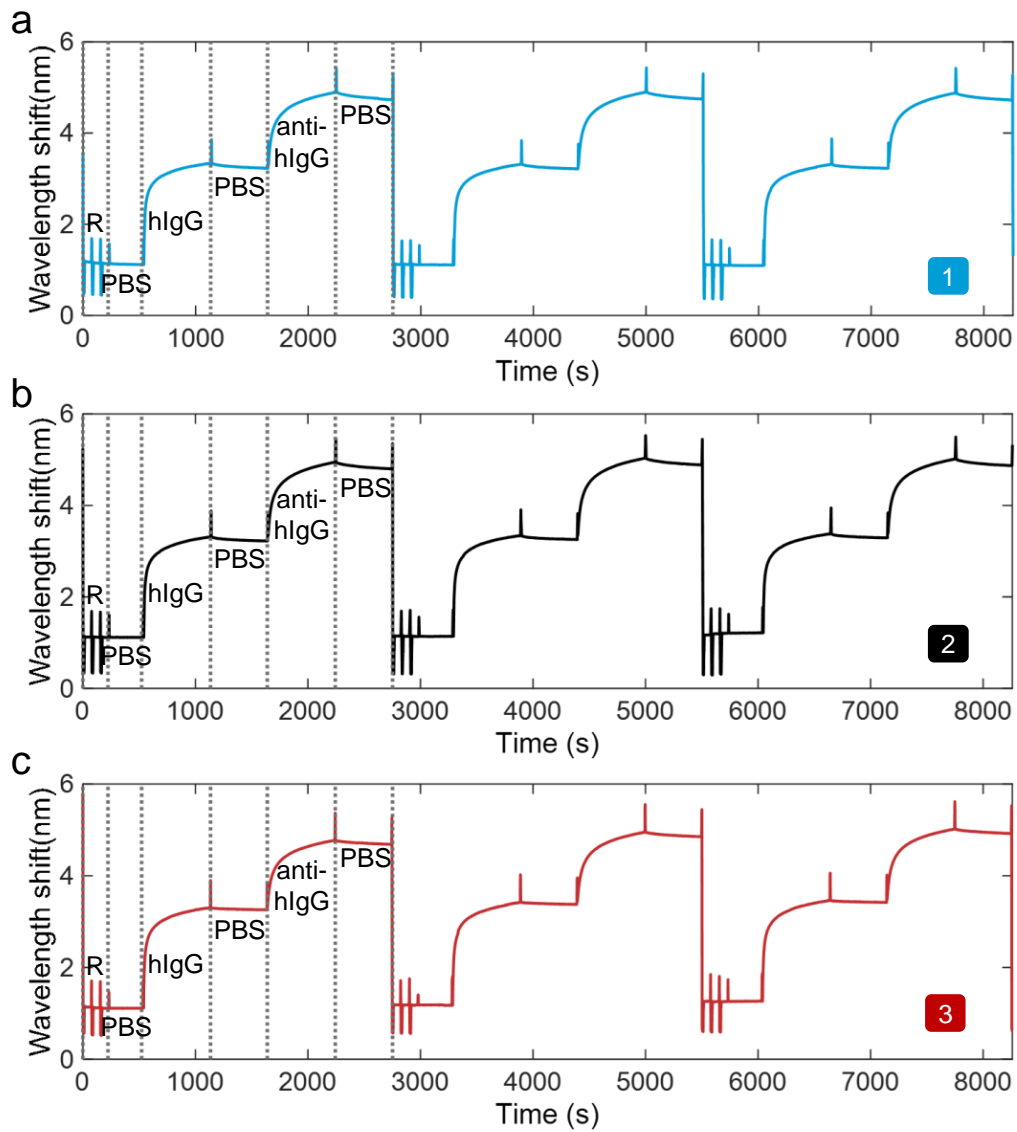

**Fig. S6. Binding and dissociation sensorgrams of hIgG and anti-hIgG.** Each row shows three repeated experiments with the same sensor device. Different rows used different sensor devices.

## References

S1. A. Checco, T. Hofmann, E. DiMasi, C. T. Black, B. M. Ocko, "Morphology of air nanobubbles trapped at hydrophobic nanopatterned surfaces," *Nano Lett.* **10**, 1354-1358 (2010).

S2. Y.-H. Lu, C.-W. Yang, C.-K. Fang, H.-C. Ko, I.-S. Hwang, "Interface-induced ordering of gas molecules

confined in a small space," *Sci. Rep.* **4**, 7189 (2014).

S3. C. Li, A.-M. Zhang, S.-P. Wang, P. Cui, "Effects of nanostructured substrates on the dynamic behavior of nanobubbles," *Phys. Rev. Fluids* **4**, 103603 (2019).

S4. A. Angulo, P. van der Linde, H. Gardeniers, M. Modestino, D. Fernández Rivas, "Influence of bubbles on the energy conversion efficiency of electrochemical reactors," *Joule* **4**, 555-579 (2020).

S5. P. G. de Gennes, "On fluid/wall slippage," *Langmuir* **18**, 3413-3414 (2002).